\documentclass[12pt]{article}
\usepackage[english]{babel}
\usepackage{amssymb,latexsym,amsmath,amscd}
\usepackage[dvips]{graphicx}

\begin{document}

\title{Standard Model gauge coupling unification}
\author
{E.K. Loginov\footnote{{\it E-mail address:} ek.loginov@mail.ru}\\
\it Ivanovo State University,\\
\it Ermaka St. 39, Ivanovo, 153025, Russia}
\maketitle

\begin{abstract}
We study the low energy evolution of coupling constants of the standard model and show that
gauge coupling unification can be achieved at the electroweak scale with a suitable
normalization. We choose the grand unification group to be the semidirect product of $Spin(8)$
by $S_3$. In this case the three low energy gauge couplings and the two scalar self-couplings
are determined in terms of two independent parameters. In particular, it gives a precise
prediction for the mass of the Higgs boson.
\end{abstract}

\section{Introduction}

The standard model (SM) is a mathematically consistent renormalizable field theory which
predicts or is consistent with all experimental facts~\cite{beri12}. It successfully predicted
the existence and form of the weak neutral current, the existence and masses of the $W$ and
$Z$ bosons, and the charm quark, as necessitated by the GIM mechanism. The charged current
weak interactions, as described by the generalized Fermi theory, were successfully
incorporated, as was quantum electrodynamics. The consistency between theory and experiment
indirectly tested the radiative corrections and ideas of renormalization and allowed the
successful prediction of the top quark mass. Nevertheless, despite the apparent striking
success of the theory, there are a lot of reasons why it is not the ultimate theory. First
there is the well-established experimental observations of neutrino oscillations which are
impossible in the SM. Secondly, some values of the SM parameters are not calculable in the
theory, notably, the fermion mass hierarchy, the hierarchy of symmetry-breaking scales, and
the Higgs boson mass. Hence the theory has far too much arbitrariness to be the final story.
Finally, there exist purely theoretical difficulties in describing hadrons by means of the
available methods of quantum field theory. These and other deficiencies of the SM motivated
the effort to construct theories with higher unification of gauge symmetries.
\par
In the framework of the grand unification hypothesis~\cite{geor74,geor74a}, it is possible to
obtain a reasonable explanation of the relation $\Lambda_{QCD}\ll M_{GUT}$ that is based on
the logarithmic renormalization-group dependence of the gauge coupling constant on the energy.
Note, however, a similar analysis is not successful for the electroweak interaction, whose
coupling constants are small at the scale $v\approx 246$ GeV. It is unrelated to any dynamical
scale and is introduced into the theory as a free parameter. One immediate consequence of the
grand unification hypothesis is a very simple explanation for the experimentally observed
charge quantization. This is because the eigenvalues of the generators of a simple non-Abelian
group are discrete while those corresponding to the Abelian group are continuous.
Unfortunately, by now LEP data have shown~\cite{land91,elli91,amal91} that simple non-SUSY
grand unifications must be excluded, initially by the increased accuracy in the measurement of
the Weinberg angle, and by early bounds on the proton lifetime~\cite{marc87}. In other words,
gauge coupling unification cannot be achieved in the SM if we choose the canonical
normalization for the SM group, i.e., the Georgi--Glashow $SU(5)$ normalization. Also, to
avoid proton decay induced by dimension-6 operators via heavy gauge boson exchanges, the gauge
coupling unification scale is constrained to be higher than about $5\times 10^{15}$ GeV.
\par
This latter restriction is not true for gauge-Higgs
unification~\cite{mant79,hoso83,hata98,kawa01}, however. In gauge-Higgs models, the
compactification scale may be of the order of the electroweak. Such unification is a very
fascinating scenario beyond the SM since the Higgs doublet is identified with the extra
component of the higher dimensional gauge field and its mass squared correction is predicted
to be finite regardless of the non-renormalizable theory. This fact has opened a new
possibility to solve the gauge hierarchy problem without, for example, supersymmetry.
Obviously, the gauge-Higgs coupling unification can be achieved at the electroweak scale only
with a suitable normalization. Note also that the unification group of the model does not
necessarily be simple. For example, such group may be represented as a product of identical
simple groups (with the same coupling constants by some discrete symmetries) or it may be
obtained as an extension of a simple Lie group by means of a finite group of operators. The
latter possibility will be considered in this paper.

\section{The $S_3\ltimes Spin(8)$ symmetry}

We begin by discussing the following simple construction. This construction arises in
connection with the following question: is it possible to embed an arbitrary group $G$ in some
group $\widetilde{G}$ with the property that every automorphism of $G$ is the restriction of
some inner automorphism of $\widetilde{G}$? Let $\Phi$ be a subgroup of $\text{Aut}\,G$. Then
for $\widetilde{G}$ one may take the set of ordered pair $\phi g$ with multiplication defined
by the rule
\begin{equation}\label{2-01}
\phi g\cdot\phi'g'=\phi\phi'g^{\phi'}g',
\end{equation}
there $\phi\in\Phi$ and $g\in G$. (We are writing pairs without their customary comma and
brackets.) The group axioms are straightforward to verify. From the rule for multiplication
(\ref{2-01}) it is immediate that $\phi^{-1}g\phi=g^{\phi}$. Hence the problem is solved. The
group $\widetilde{G}$ is called the extension of the group $G$ by means of the automorphisms
in $\Phi$ and denoted as $\Phi\ltimes G$. Alternatively one says that $\widetilde{G}$ is a
semidirect product of $G$ by $\Phi$.
\par
Now let $\Phi$ be a subgroup of the outer automorphisms group of $G$. Suppose $V$ is a
representation space of $G$. Then the representation of $G$ in $V$ induces a representation of
$\widetilde{G}$ in the direct sum $\widetilde{V}=V_1\oplus\dots\oplus V_{n}$, where each
direct summand $V_{i}$ is isomorphic to $V$ and $n=|\Phi|$. For the space $V_{i}$ one may take
the ordered pair $\phi_{i}V$, where $\phi_{i}\in\Phi$. Then the action of $\widetilde{G}$ on
$\widetilde{V}$ can be written as
\begin{equation}\label{2-02}
\phi g\cdot\oplus_{i}\phi_{i}V=\oplus_{i}\phi\phi_{i}g^{\phi_{i}}V.
\end{equation}
Let $G_0$ be a set of elements of $G$ such that $g^{\phi}=g$ for all $\phi\in\Phi$. Clearly,
it is a $\Phi$-invariant subgroup of $G$. Using the formula (\ref{2-02}) one may define a
representation of $G_0$ on $\widetilde{V}$. It is easy to prove that $G_0$ acts equivalently
on each direct summand of $\widetilde{V}$.
\par
Let $G$ be a simple gauge group. If the normalization of the generators of $G$ are fixed, then
the gauge couplings will be the same for both $G$ and $G_0$. Suppose that for the energy scale
$\mu>M_0$ an $G$ gauge theory possesses both discrete and gauge symmetries, whereas for
$\mu=M_0$ the symmetries breaking $\widetilde{G}\to G_0$ to take place. Then the
representation space of $G_0$ reduces to $V$ and hence the normalization of the generators of
$G_0$ must be changed. Since the definition of coupling constants depends on the normalization
of the generators it follows that the gauge coupling of $G_0$ should be also change, namely
$g^2\to g^2|\Phi|$ as $\widetilde{V}\to V$.
\par
Now we suppose $G=Spin(8)$. This group has the outer automorphisms group
$S_3=\langle\rho,\sigma\rangle$, where $\rho^3=\sigma^2=(\rho\sigma)^2=1$, and two
Majorana--Weyl real eght-dimensional representations that related to the eght-dimensional real
vector representation by the action of $S_3$. The group $Spin(8)$ cannot contain the SM group
as a subgroup, but it contains disjoint subgroups $G_3$ and $G_2\times G_1$ that are
isomorphic to $SU(3)$ and $SU(2)\times U(1)$, respectively. Moreover, we always can choose
these subgroups in the following manner:
\medskip\par
(i) $g^{\phi}\ne g$ for $1\ne\phi\in S_3$ and $g\in G_1$,
\par
(ii) $g^{\phi}\ne g=g^{\rho\sigma}$ for $1\ne\phi\ne\rho\sigma$ and $g\in G_2$,
\par
(iii) $g^{\phi}=g$ for all $\phi\in S_3$ and $g\in G_3$.
\par\medskip\noindent
Hence, if for the energy scale $\mu=M_0$ the discrete and gauge symmetries breaking of
$S_3\ltimes Spin(8)$ to take place, then the gauge couplings of $G_3$, $G_2$, and $G_1$ should
be satisfy
\begin{equation}\label{2-05}
g_3=\sqrt{3}g_2=\sqrt{6}g_1
\end{equation}
as $\mu=M_0$.
\par
Just as for $Spin(8)$, the group $S_3\ltimes Spin(8)$ cannot contain the SM group as a
subgroup. Nevertheless, in the next section we shall show that there is a way to break the
symmetry by
\begin{equation}\label{2-04}
S_3\ltimes Spin(8)^{\begin{CD}@>{SU(2)}>>\end{CD}} SU(3)\times U(1).
\end{equation}
Moreover, the symmetry breaking is such that the gauge couplings of the SM group also satisfy
(\ref{2-05}). This possibility of the breaking is based on the following mathematical
construction~\cite{doro78}.
\par
As was remarked before, the group $G$ admits the outer automorphisms $\rho$ and $\sigma$. Let
\begin{equation}
G^{\sigma}=\{g\in G\mid g^{\sigma}=g\}
\end{equation}
(i.e. $G^{\sigma}$ is the centralizer of $\sigma$ in $G$). We denote by $S$ the factor space
$G/G^{\sigma}$. Our nearest aim is to define a binary composition on the cosets of
$G^{\sigma}$. Define
\begin{equation}
L=\{g^{-1}g^{\rho\sigma}\mid g\in G\}.
\end{equation}
For each left coset $gG^{\sigma}$, there exists exactly one element $L_{a}$ in $L$ such that
$L\cap gG^{\sigma}=\{L_{a}\}$. This defines a permutation representation of $G$ on $S$; if
$gL_{a}G^{\sigma}=L_{b}G^{\sigma}$ for $g\in G$ let $gL_{a}=L_{b}$. Using this permutation
representation, we may define a binary composition in $S$. If
$L_{a}L_{b}G^{\sigma}=L_{c}G^{\sigma}$, define $ab=c$. This binary composition make $S$ into a
nonassociative loop isomorphic to $\mathbb{S}^7$, which is defined in Appendix A. In
particular, this defines the permutation representation of $G$ on $\mathbb{S}^7$. Extending
this action by linearity on $\mathbb{O}$, we obtain the eight-dimensional spinor
representation of $G$.
\par
Apart from the left action of $G$ on $S$, there exists the trivial right action of $G$ on $S$
with $G^{\sigma}$ acting on itself by multiplication on the right. Suppose $G_3\subset
G^{\sigma}$ and $(G_2\times G_1)\cap G^{\sigma}=1$. Then the (left ant right) actions of $G$
on $S$ induce the left action of $G_2\times G_1$ on $S$ and the right action of $G_3$ on
$G^{\sigma}$. Obviously, these group actions are independent of each other.

\section{Gauge-Higgs unification}

In this section we briefly discuss the gauge-Higgs model based on the group $S_3\ltimes
Spin(8)$. Consider the $Spin(8)$-invariant gauge theory defined on the manifold
$M=M_{3,1}\times\mathbb{S}^7$, where $M_{3,1}$ is the Minkowski spacetime and $\mathbb{S}^7$
is the seven-dimensional sphere. (We assume that this sphere is equipped with the octonionic
multiplication.) Suppose that this theory possess a symmetry under a discrete group $K$ of
inner automorphisms. Further, let $A(x,y)$ and $C(x,y)$ be all gauge fields in the theory and
only the fields $C(x,y)\in so(7)_{v}$, where $so(7)_{v}$ is the Lie algebra of $G^{\sigma}$.
Following~\cite{hebe01}, we declare that only field configurations invariant under the action
\begin{equation}\label{2-06}
K:\left\{\aligned A(x,y)&\to M(k) A(x,k^{-1}[y])\\
C(x,y)&\to N(k)C(x,k^{-1}[y])\endaligned\right.,
\end{equation}
are physical. Here $M(k)$ and $N(k)$ are matrix representations of $K$ and $k[y]$ is the image
of the point $y\in\mathbb{S}^7$ under the operation of $k\in K$. But unlike the standard
orbifolding conditions, we suppose that $M(K)\ne N(K)$.
\par
In order to determine automorphisms that are responsible for the symmetry breaking
(\ref{2-04}), we must first select the group $K$. Suppose $K=\mathbb{Z}_4\times\mathbb{Z}_2$.
Further, suppose that the subgroup $H$ in $Spin(8)$ is generated by the operators $I=R_7$ and
$J=L_7^{-1}L_5L_2$, which are defined in Appendix B. Obviously, $IJ=JI$ and $I^4=J^2=1$. It
follows from this that $H=H_{I}\times H_{J}$, where the subgroups $H_{I}$ and $H_{J}$ are
generated by $I$ and $J$, respectively, i.e., $H\simeq\mathbb{Z}_4\times\mathbb{Z}_2$. Define
the action of $K$ on $\mathbb{S}^7$ and the representations $K\to M(K)$ and $K\to N(K)$ as
follows. Let $f$, $f_{I}$, and $f_{J}$ be homomorphisms of $K$ onto $H$, $H_{I}$, and $H_{J}$
respectively. Then
\begin{align}
M(k)A(x,y)&=f(k)^{-1}A(x,y)f(k),\\
N(k)C(x,y)&=f_{I}(k)^{-1}C(x,y)f_{I}(k),\\
k^{-1}[y]&=f_{J}(k)^{-1}y.
\end{align}
\par
We now focus our attention on the symmetry breaking at the fixed points. Since the factor
group $K/K_{J}$ acts on $\mathbb{S}^7$ trivially, it follows that the $Spin(8)$ gauge symmetry
is reduced (under the given action) to the centralizer of $I$ in $Spin(8)$, i.e., to
$SU(4)_{s}\times U(1)$ (see also Appendix C). Therefore the fields $A(x,y)$ and $C(x,y)$ take
its values in the Lie algebra $su(4)_{s}\oplus u(1)$. In particular, $C(x,y)$ takes its values
in $su(3)_{s}$ (which is the intersection of $su(4)_{s}$ and $so(7)_{v}$). Now consider the
$SU(4)_{s}\times U(1)$ symmetry breaking by $K_{J}$. It follows from (\ref{2-04}) that the
$SU(3)_{s}$ symmetry must be preserved under the action of $K_{J}$ on $\mathbb{S}^7$. This is
possible only if the components of $C(x,y)$ are independent of $y$, i.e., if
\begin{equation}
C_{M}(x,y)=ig_{s}C^{p}_{\mu}(x)\frac{\lambda_{p}}{2},
\end{equation}
where $\lambda_{p}$ are the usual Gell-Mann matrices for $SU(3)$ (cf. the formulas in Appendix
C).
\par
On the contrary, the unbroken symmetries of $A(x,y)$ must belong to the centralizer of $J$ in
$SU(4)_{s}\times U(1)$. Using (\ref{b-03}) and the explicit form of the generators of
$SU(4)_{s}\times U(1)$, we prove that the fields $A(x,y)$ take its values in the Lie algebra
$su(2)_{s}\oplus u(1)$. Hence we can define these fields by
\begin{align}
A_{M}(x)&=igA^{k}_{\mu}(x)\frac{\sigma_{k}}{2},\\
B_{M}(x,y)&=\{ig'B_{\mu}(x),g'\phi_{a}(x)\},\label{2-07}
\end{align}
where $\sigma_{k}$ are the standard Pauli matrices and $\phi_{a}$ are the components of
$\phi(x)\in\mathbb{O}$. Strictly speaking, we would have to write $\phi(x)\in\mathbb{S}^7$.
However, in this case the field $\phi(x)$ will be unobservable. The point is that for the
Kaluza--Klein type theories to be able to describe the observed four-dimensional world it is
necessary for the extra spatial dimensions to be compactified down to a size which we do not
probe in particle physics experiments (e.g. the Planck length). Therefore we suppose that the
field $\phi(x)$ has quantum fluctuations such that $\phi(x)\in\mathbb{O}$ but not
$\mathbb{S}^7$. In this case exactly one component of $\phi(x)$  (scalar field) will be
observed.
\par
We now consider the action of $K_{J}$ on $\mathbb{S}^7$. It is easily shown that the condition
$Jy=y$ is equivalent to $(e_5,e_2,y)=0$. Hence $y$ belong to an associative subalgebra of
$\mathbb{O}$ generated by $e_5$ and $e_2$. Obviously, this is the algebra of quaternions with
the basis $1,e_2,e_5,e_7$. The complexification of $\mathbb{O}$ transform this subalgebra into
a two-dimensional unitary space. (Denote it by the symbol $\varPhi$.) We have proved, in fact,
that all nonzero fields $\phi(x)$ in (\ref{2-07}) must belong to $\varPhi$. Thus, if we
identify $\phi(x)$ as the massless Higgs doublet, then we obtain a complete set of boson
fields of the SM.
\par
Finally, consider the SM Lagrangian for the field $\phi(x)$
\begin{equation}\label{2-08}
{\cal L}_{H}=(D_{\mu}\phi)^{\dag}(D^{\mu}\phi)+\mu^2\phi^2-\lambda\phi^4,
\end{equation}
and suppose that the fields $B_{\mu}(x)$ and $\phi_{a}(x)$ in (\ref{2-08}) are components of
$B(x,y)$. Then we have
\begin{equation}\label{2-09}
\lambda=g'^2.
\end{equation}
As usual the Higgs vacuum expectation value breaks the $SU(2)\times U(1)$ symmetry down to
$U(1)$. As a result we have the symmetry breaking (\ref{2-04}).

\section{Gauge coupling unification}

The conditions (\ref{2-05}) are valid in the $S_3\ltimes Spin(8)$ limit. Now we need to study
the regime $\mu<M_0$. The evolution of the SM gauge coupling constants in the one-loop
approximation is controlled by the renormalization group equation
\begin{equation}\label{3-01}
\frac{d\alpha_{n}^{-1}(\mu)}{d\ln\mu}=\frac{b_{n}}{6\pi},
\end{equation}
where $b_1=-2n_1$, $b_2=22-2n_2$, $b_3=33-2n_3$, and $\alpha_{n}=g^2_{n}/4\pi$. (We have
ignored the contribution coming from the Higgs scalar and higher-order effects.) It follows
from (\ref{2-05}) that the generators of SM group in the fundamental representation should be
normalize by the condition $6n_3=2n_2=n_1=N_{f}$, where $N_{f}$ is the number of quark
flavors. Expressing the low-energy couplings in terms of more familiar parameters, we can
represent the solutions of Eq. (\ref{3-01}) as
\begin{align}
\alpha_{s}^{-1}(\mu)
&=\alpha_3^{-1}(M_0)-\frac{b_3}{6\pi}\ln\frac{M_0}{\mu},\label{3-02}\\
\alpha^{-1}(\mu)\sin^2\theta_{\mu}
&=\alpha_2^{-1}(M_0)-\frac{b_2}{6\pi}\ln\frac{M_0}{\mu},\label{3-03}\\
\frac35\alpha^{-1}(\mu)\cos^2\theta_{\mu}
&=\alpha_1^{-1}(M_0)-\frac{b_1}{6\pi}\ln\frac{M_0}{\mu}.\label{3-04}
\end{align}
Taking the linear combination [$12\times\text{Eq.}\,(\ref{3-02})
-18\times\text{Eq.}\,(\ref{3-03})+7\times\text{Eq.}\,(\ref{3-04})$] and using the relations
(\ref{2-05}), we have
\begin{equation}\label{3-05}
\sin^2\theta_{\mu}=\frac{7}{37}+\frac{60}{111}\frac{\alpha(\mu)}{\alpha_{s}(\mu)}.
\end{equation}
Obviously, Eq. (\ref{3-05}) implies a non-trivial consistency condition among the gauge
couplings. Taking the linear combination [$-78\times\text{Eq.}\,(\ref{3-02})
+6\times\text{Eq.}\,(\ref{3-03})+10\times\text{Eq.}\,(\ref{3-04})$] and again using the
relations (\ref{2-05}), we have
\begin{equation}\label{3-06}
\ln\frac{M_0}{\mu}=\frac{6\pi}{407}\left[\alpha^{-1}(\mu)-13\alpha_{s}^{-1}(\mu)\right].
\end{equation}
This determines the unification scale $M_0$. Also, combining Eqs. (\ref{3-05}) and
(\ref{3-06}), we obtain
\begin{equation}\label{3-07}
\sin^2\theta_{\mu}=\frac{3}{13}-\frac{110\alpha(\mu)}{39\pi}\ln\frac{M_0}{\mu}.
\end{equation}
Finally, it follows easily from Eqs. (\ref{3-02})--(\ref{3-04}) that the running electroweak
and strong gauge coupling constants satisfy
\begin{align}
\alpha^{-1}(\mu)&=\alpha^{-1}(M_0)-\frac{66-13N_{f}}{18\pi}\ln\frac{M_0}{\mu},\label{3-08}\\
\alpha^{-1}_{s}(\mu)
&=\alpha_{s}^{-1}(M_0)-\frac{99-N_{f}}{18\pi}\ln\frac{M_0}{\mu},\label{3-09}
\end{align}
where the gauge couplings are connected by the relation $\alpha_{s}(M_0)=13\alpha(M_0)$. Note
that the choice of normalization of the generators essentially influences on the behavior of
the gauge couplings by changing its values in fixed points. So for example Eqs. (\ref{3-08})
and (\ref{3-09}) will differ from that obtained in the SM. Therefore we must have a rule which
permits to compare the gauge coupling constants in our (non-canonical) and the canonical
normalizations. We will extract this rule from the renormalization on-shell scheme.
\par
The on-shell scheme~\cite{sirl80,sirl84,kenn89,kenn89a,bard89,holl90} (see also
Ref.~\cite{lang95}) promotes the tree-level formula $\sin^2\theta_{W}=1-M^2_{W}/M^2_{Z}$ to a
definition of the renormalized $\sin^2\theta_{W}$ to all orders in perturbation theory, i.e.,
\begin{equation}\label{3-10}
\sin^2\theta_{W}=\frac{\pi\alpha v^2}{M^2_{W}(1-\Delta r)},
\end{equation}
where $\Delta r$ summarizes the higher order terms. Here $\alpha$ is the fine structure
constant, $M_{W}$ is the mass of the charged gauge boson, and $v=(\sqrt{2}G_{F})^{-1/2}$ is
the vacuum expectation value. One finds $\Delta r=\Delta r_0-\Delta r'$, where $\Delta
r_0=1-\alpha/\alpha(M_{Z})$ is due to the running of $\alpha$ and $\Delta r'$ represents the
top quark mass $m_{t}$ and the Higgs boson mass $M_{H}$ dependence. Using the formal expansion
$(1+\Delta r')^{-1}=1-\Delta r'+\dots$, we can rewrite the formula (\ref{3-10}) in the form
\begin{equation}\label{3-11}
\sin^2\theta_{W}=\frac{\pi\alpha(M_{Z})v^2}{M^2_{W}}\left(1-\frac{\alpha(M_{Z})}{\alpha}\Delta
r'+\dots\right).
\end{equation}
In the on-shell scheme the value of $\sin^2\theta_{W}$ is independent of the normalization of
the generators. We suppose that the value of $\alpha(\mu)$ in the fixed point $\mu=M_{Z}$ is
the same for both the canonical and non-canonical normalizations. Using this condition, we can
now compare values of the running coupling constant $\alpha(\mu)$ in the two normalizations.
\par
\begin{figure}[htb]
\centering
\includegraphics[scale=0.4,angle=270]{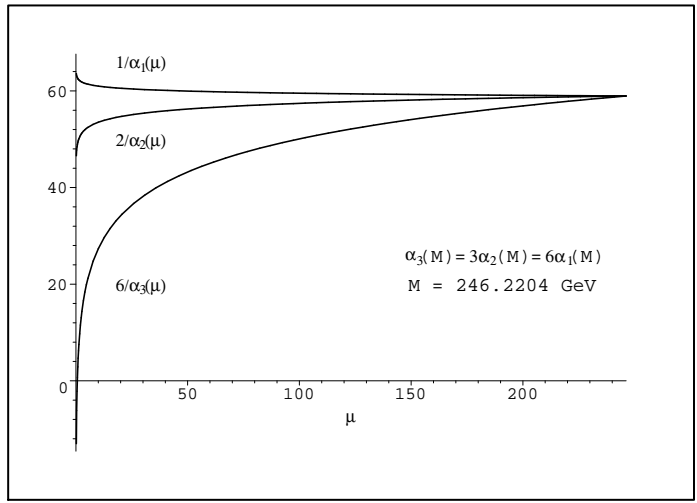}
\caption{One-loop gauge coupling unification for the SM with the non-canonical normalization.}
\label{fig:one-loop}
\end{figure}

Following Ref.~\cite{pesk90,alta91,novi93}, we remove the $(m_{t},M_{H})$ dependent term from
$\Delta r$ and write the renormalized $\sin^2\theta_{\mu}$ (in the non-canonical
normalizations) as
\begin{equation}\label{3-12}
\sin^2\theta_{\mu}=\frac{\pi\alpha(\mu') v^2}{M^2_{W}}
\end{equation}
for $M_{Z}\leq\mu\leq\mu'\leq v$. Further, we suppose that the unification scale coincides
with the electroweak scale (i.e., $M_0=v$) and that the left-hand side of Eqs. (\ref{3-07})
and (\ref{3-12}) are equal as $\mu=M_0$. In this case
\begin{equation}\label{3-13}
\frac{M_{W}}{M_0}=\sqrt{\frac{13\pi\alpha_0}{3}},
\end{equation}
where $\alpha_0=\alpha(M_0)$. It follows from (\ref{3-05})--(\ref{3-09}) and (\ref{3-13}) that
the three parameters $\alpha(M_{Z})$, $\alpha_{s}(M_{Z})$, and $\sin\theta_{W}$ of the SM are
now determined in terms of one independent parameters $\alpha_0$. We show the gauge coupling
unification in Fig.~\ref{fig:one-loop}. Thus, there are two predictions.

In conclusion, we show that the values of the coupling constants and the masses of the gauge
bosons which are deduced from the SM are compatible with these predictions. Using
$\alpha_0^{-1}=127.726$ and $N_{f}=6$ yield $\alpha^{-1}(M_{Z})=127.937$,
$\alpha_{s}(M_{Z})=0.1221$, and $\sin^2\theta_{W}=0.2229$. With more careful treatment of
two-order effects, one obtains $\alpha_{s}(M_{Z})=0.1210$. (Other parameters are changed
unessentially.) These values are compatible with the SM predictions in
Refs.~\cite{scha06,baik08,mart09,beth09}. (For a recent review, see Ref.~\cite{beri12} and
references therein.) This means, in particular that the value of $\alpha_{s}(M_{Z})$ may also
be chosen the same for both the canonical and non-canonical normalizations. Using
$M_0=246.2204$ GeV yield $M_{W}=80.3841$ GeV and $M_{Z}=91.1876$ GeV. This is also compatible
with the SM predictions.
\par

\section{Higgs boson mass}

The condition (\ref{2-09}) gives a precise prediction about the Higgs mass $m_{\phi}$. Here we
follow the presentation of Coleman and Weinberg~\cite{cole73} (see also~\cite{chen84}). The
one-loop effective potential of $SU(2)\times U(1)$ gauge theory is given by
\begin{equation}\label{3-14}
V(\phi)=-\mu^2\phi^2+\lambda\phi^4+C\phi^4\ln\frac{\phi^2}{M^2},
\end{equation}
where $M$ is an arbitrary mass parameter and
\begin{equation}
C=\frac{1}{16\pi v^4}\left(3\sum_{b}m_{b}^4+m_{\phi}^4-4\sum_{f}m_{f}^4\right).
\end{equation}
Here the indices $b$ and $f$ run over the vector bosons and fermions (the top quark
contribution is excluded), and the mass of the Higgs scalar is taken to zeroth order, i.e.,
\begin{equation}\label{3-15}
m^2_{\phi}=2\mu^2=2\lambda v^2.
\end{equation}
With (\ref{3-14}), we can obtain the mass of the Higgs particle. It is given by
\begin{equation}
m^2_{\phi}=2v^2\left[\lambda+C\left(\ln\frac{v^2}{2M^2}+\frac32\right)\right].
\end{equation}
\par
We now use the condition (\ref{2-09}). It follows from the formulas (\ref{2-05}) and
(\ref{3-13}) that
\begin{equation}
\lambda=\frac{3}{10}g^2=\frac{26\pi\alpha_0}{5}.
\end{equation}
Substituting this expression into (\ref{3-15}), one finds the mass of the Higgs scalar to
zeroth order ($m_{\phi}=124.53$ GeV). Knowing the masses of the vector bosons and fermions,
one may calculate $C$. (We obtain $C=0.0012$.) Finally, choosing the mass parameter $M=M_{Z}$,
we have the Higgs boson mass $m_{\phi}=126.15$ GeV. This is agrees quite well with the
experimental results that were recently obtained in~\cite{aad12,chat12}.

\section{Conclusion}

In this paper, we have shown that the gauge-Higgs model based on the group $S_3\ltimes
Spin(8)$ can be considered as a candidate for the real physical theory. It do not contradict
SM (at least at the bosonic sector) and gives precise predictions for the masses of the gauge
bosons and the Higgs scalar. Here we make two general remarks.
\par
The group $Spin(8)$ occupies a special position among the simple Lie groups since only it has
outer automorphism group $S_3$. Namely this property of $Spin(8)$ permits to define the
non-canonical normalization in a natural way and to get the gauge coupling constants
unification. To describe the $S_3$-symmetry breaking, we embedded the $Spin(8)$-gauge theory
in the model with larger global symmetry groups. The motivation for this is that whatever the
high energy physics producing the spontaneous breaking of the gauge group, it is likely to
possess a larger global symmetry than the gauged one. We risk to suppose that the existence of
discrete $S_3$-symmetry is related to a duplication of the fermionic structure in the SM. We
suppose that $S_3\ltimes K$ is the discrete flavor symmetry group. But of course this is only
a hypothesis.
\par
In order to describe the gauge-Higgs unification, we have used the relation between $Spin(8)$
and the algebra of octonions. Generally speaking, this approach is not new. Properties of
octonions was used earlier to describe various mechanisms of compactification of $d=11$
supergravity down to $d=4$~\cite{engl82} (see also the review~\cite{duff86}) and to find
solutions of the low-energy heterotic string theory~\cite{harv91}. There have been many other
attempts over the years to incorporate this algebra into physics. The present paper is a next
step in this direction.

\numberwithin{equation}{section}
\appendix
\section{Octonions}

We recall that the algebra of octonions $\mathbb O$ is a real linear algebra with the
canonical basis $e_0=1,e_{1},\dots,e_{7}$ such that
\begin{equation}\label{a-01}
e_{i}e_{j}=-\delta_{ij}+c_{ijk}e_{k},
\end{equation}
where the structure constants $c_{ijk}$ are completely antisymmetric and nonzero and equal to
unity for the seven combinations (or cycles)
\begin{equation}
(ijk)=(123),(154),(167),(264),(275),(347),(365).
\end{equation}
The algebra of octonions is not associative but alternative, i.e. the associator
\begin{equation}\label{a-02}
(x,y,z)=(xy)z-x(yz)
\end{equation}
is totally antisymmetric in $x,y,z$. Consequently, any two elements of $\mathbb O$ generate an
associative subalgebra. The algebra $\mathbb O$ permits the involution (anti-automorphism of
period two) $x\to\bar x$ such that the elements
\begin{equation}
t(x)=x+\bar x,\qquad n(x)=\bar xx
\end{equation}
are in $\mathbb R$. In the canonical basis, this involution is defined by $\bar e_{i}=-e_{i}$.
It follows that the bilinear form
\begin{equation}
(x,y)=\frac12(\bar xy+\bar yx)
\end{equation}
is positive definite and defines an inner product on $\mathbb O$. It is easy to prove that the
quadratic form $n(x)$ permits the composition
\begin{equation}
n(xy)=n(x)n(y).
\end{equation}
It follows from this that the seven-dimensional sphere
\begin{equation}
\mathbb{S}^7=\{x\in\mathbb O\mid n(x)=1\}
\end{equation}
is closed relative to the multiplication in $\mathbb{O}$. Finally, since the quadratic form
$n(x)$ is positive definite, it follows that $\mathbb O$ is a division algebra.

\section{Triality}

Let $x$ be any element of $\mathbb{O}$. The left multiplications $L_{x}$ and right
multiplications $R_{x}$ of $\mathbb{O}$ which are determined by $x$ are defined by
\begin{equation}
L_{x}y=xy,\qquad R_{x}y=yx
\end{equation}
for all $y$ in $\mathbb{O}$. Clearly $L_{x}$ and $R_{x}$ are linear operators on $\mathbb{O}$.
We choose the canonical basis and denote by $L_{i}$ and $R_{i}$ the operators $L_{e_{i}}$ and
$R_{e_{i}}$ respectively. Then from (\ref{a-01}) and the fully antisymmetry of the associator
(\ref{a-02}), we get
\begin{equation}\label{b-03}
L_{i}L_{j}+L_{j}L_{i}=-2\delta_{ij}I,
\end{equation}
where $I$ is the identity $8\times 8$ matrix. (Of course, a similar formula is true for the
right multiplications.) Hence $L_1,\dots,L_7$ are generators of the Clifford algebra
$Cl_{0,7}(\mathbb R)$, and therefore they generate the Lie algebra $so(8)$. This is the Lie
multiplication algebra of $\mathbb{O}$.
\par
In this algebra, we separate the subspaces $L$ spanned by the operators $L_{i}$ and the
subalgebra $so(7)_{s}$ spanned by the operators
\begin{align}
S_{i}&=L_{i}+2R_{i},\\
D_{ij}&=L_{[e_{i},e_{j}]}-R_{[e_{i},e_{j}]}-3[L_{i},R_{j}].
\end{align}
(The latter linearly generate the 14-dimensional exceptional simple Lie algebra $g_2$.) This
imply that the algebra $so(8)$ decomposes into the direct sum
\begin{equation}
so(8)=so(7)_{s}\oplus L.
\end{equation}
The algebra $so(8)$ admits the outer automorphisms $\rho$ and $\sigma$ of orders 3 and 2
respectively. We may define them by
\begin{equation}\label{b-01}
\left.\aligned L_{i}^{\rho}&=R_{i},&\quad R_{i}^{\rho}&=-L_{i}-R_{i},\\
L_{i}^{\sigma}&=-R_{i},&\quad R_{i}^{\sigma}&=-L_{i}.\endaligned\right\}
\end{equation}
Obviously, the automorphisms $\rho\sigma$, $\sigma\rho$, and $\sigma$ fixe all elements of
$so(7)_{s}$, $so(7)_{c}=so(7)_{s}^{\rho}$, and $so(7)_{v}=so(7)_{s}^{\rho^2}$, respectively.
The elements of intersection of the subalgebras, i.e. the elements of $g_2$, is fixed by
$\rho$.
\par
Just as for $so(8)$, the group $Spin(8)$ also admits the outer automorphisms $\rho$ and
$\sigma$. According to (\ref{b-01}), they are defined by
\begin{equation}
\left.\aligned L_{a}^{\rho}&=R_{a},&\quad R_{a}^{\rho}&=L_{a}^{-1}R_{a}^{-1},\\
L_{a}^{\sigma}&=R_{a}^{-1},&\quad R_{a}^{\sigma}&=L_{a}^{-1},\endaligned\right\}
\end{equation}
where $a\in\mathbb{S}^7$. The automorphisms $\rho\sigma$, $\sigma\rho$, and $\sigma$ fixe the
elements of $SO(7)_{s}$, $SO(7)_{c}=SO(7)_{s}^{\rho}$, and $SO(7)_{v}=SO(7)_{s}^{\rho^2}$
respectively. The intersection of the group, i.e. $G_2$, is fixed by $\rho$. Here $SO(7)_{v}$
is generated by the elements
\begin{align}
R_{a}^{-1}L_{a}&=L_{a}R_{a}^{-1},\\
L_{ab}^{-1}L_{a}L_{b}&=R_{ab}R_{a}^{-1}R_{b}^{-1}.\label{b-02}
\end{align}
Note also that these automorphisms permute inequivalent irreducible representations
$\bf{8}_{s}$, $\bf{8}_{c}$, and $\bf{8}_{v}$ of the $Spin(8)$ group having the same
dimensionality.

\section{Complexification}

Suppose $\mathbb{C}$ is a subalgebra of $\mathbb{O}$ spanned by the elements $1$ and $i=e_7$.
We may consider $\mathbb{O}$ as a four dimensional complex (or rather unitary) space relative
to the multiplication $ax$, where $a\in\mathbb{C}$ and $x\in\mathbb{O}$. This space is
invariant under the unitary group, $SU(4)_{s}\times U(1)$, the Lie algebra of which decomposes
into the direct sum
\begin{equation}
su(4)_{s}\oplus u(1)=su(3)_{s}\oplus su(2)_{s}\oplus u(1)\oplus V_{s}
\end{equation}
of the subspaces (but not the Lie subalgebras). We write down the generators of
$SU(4)_{s}\times U(1)$ in the explicit form.
\bigskip

1) $\mathbf{su(3)_{s}}$:
\begin{align}
D_{53}-D_{42}&=6(e_{21}-e_{12})\\
D_{15}-D_{26}&=6(e_{13}-e_{31})\\
D_{31}-D_{64}&=6(e_{32}-e_{23})\\
D_{23}-D_{54}&=6i(e_{21}+e_{12})\\
D_{65}-D_{12}&=6i(e_{13}+e_{31})\\
D_{41}-D_{36}&=6i(e_{32}+e_{23})\\
D_{16}-D_{34}&=6i(e_{22}-e_{33})\\
D_{52}&=2i(e_{22}+e_{33}-2e_{11})
\end{align}
2) $\mathbf{su(2)_{s}\oplus u(1)}$:
\begin{align}
L_2+L_7L_5&=2(e_{10}-e_{01})\\
L_5+L_2L_7&=2i(e_{01}+e_{10})\\
L_7+L_5L_2&=2i(e_{00}-e_{11})\\
R_7&=i(e_{00}+e_{11}+e_{22}+e_{33})
\end{align}
3) $\mathbf{V_{s}}$:
\begin{align}
L_4+L_7L_3&=2(e_{20}-e_{02})\\
L_6+L_7L_1&=2(e_{30}-e_{03})\\
L_3+L_4L_7&=2i(e_{02}+e_{20})\\
L_1+L_6L_7&=2i(e_{03}+e_{30})
\end{align}
Here $e_{ij}$ is the $4\times4$ matrix with $(i,j)$th entry 1, and all other entries 0. Note
that the automorphisms $\rho$ and $\sigma$ fixe the elements of $su(3)_{s}$ and the
automorphism $\rho\sigma$ fixes the elements of $su(2)_{s}$ and $V_{s}$, while $R_7\in u(1)$
is not invariant under any element of $S_3$. Note also that $SU(4)_{s}\times U(1)$ is the
centralizer of $R_7$ in $Spin(8)$.

\end{document}